\RequirePackage{fix-cm}
\documentclass[twocolumn]{svjour3}          
\smartqed  
\usepackage{graphicx}
\usepackage{amsfonts}
\usepackage{color}
\begin{document}

\title{Absence of `fragility' and mechanical response of jammed granular materials
\thanks{This paper is dedicated to the memory of \textit {Isaac Goldhirsch}, a great scientist and a wonderful person.}
}


\author{Raffaele Pastore        \and
        Massimo Pica Ciamarra 
        \and Antonio Coniglio
}


\institute{R. Pastore \and M. Pica Ciamarra \and A. Coniglio \at
               CNR-SPIN, Dip. Scienze Fisiche, Universit\'a di Napoli "Federico II", 
               Naples, Italy. \\
             \email{raffaele.pastore@spin.cnr.it}\\
} 
         

\date{Received: date / Accepted: date}

\maketitle

\begin{abstract}
We perform molecular dynamic (MD) simulations of frictional non-thermal
particles driven by an externally applied shear stress.
After the system jams following a transient flow, 
we probe its mechanical response in order 
to clarify whether the resulting solid is `fragile'. 
We find the system to respond elastically and isotropically
to small perturbations of the shear stress, 
suggesting absence of fragility. These results are interpreted in terms
of the energy landscape of dissipative systems.
For the same values of the control parameters, we check the behaviour of the system during a stress cycle.
Increasing the  maximum stress value, a crossover from a visco-elastic to a plastic regime is observed.

\end{abstract}

\section{Introduction}
\label{intro}
Among the peculiarities of granular systems~\cite{Goldhirsch1}, there are jamming processes
consisting in a sudden transition from a flowing to a solid
state \cite{nagel,o'hern}. In particular, if submitted to an anisotropic stress, 
a granular material can jam after a transient flow.
This transition is observed neither at low densities,
where granular materials behave like fluids, nor at very high densities, where
they behave as disordered solids. 
The jammed solids obtained under these conditions, are so unconventional that it has been claimed
to represent an absolutely new state of matter, 
which have been termed `fragile matter'~\cite{Cates1}.
Fragility is based on the idea that during the flow particles rearrange until they find 
a configuration able to balance the external stress. 
It is speculated that the force chains sustaining this configuration are strictly related to the
flow-generating stress, so that they can support incremental stress oriented as the former (compatible stress). 
By contrast, the may not sustain a stress applied in a different direction (incompatible stress), which may
allow the system to restart flowing.
This resembles the case of a pile of sands obtained in the presence gravity. 
The pile responds as a solids to stress directed as the weight, but it falls down under
the action, for example, of a shear stress.

A more precise statement of the idea of fragility has been provided in the limit of hard spheres \cite{Cates1,Cates2,Cates3}. 
Consider, for example, a system jammed in the presence of a constant shear stress $\sigma_{zx}$:
in this framework compatible stress
variations, $\sigma'= a \sigma_{zx}$ with $a > 1$, lead to an elastic response.
Conversely incompatible stresses, such as $\sigma'= a \sigma_{zx}+b\sigma_{zy}$ 
restore flow, also for infinitesimal values of $b$. 
In this sense, these systems are `fragile' and strongly
differ from any ordinary visco-elastic or elasto-plastic material.


Previous results suggest that, in the presence of a shear stress, a fragile behavior can only be observed in 
frictional granular systems 
\cite{PCC09,o'hern06,Zhang,Makse,van
Hecke,Magnanimo,Behringer,Grebenkow,Sandnes,PRE2011}. 
Indeed, as reported in Refs.~\cite{PCC09,o'hern06}, flowing frictionless systems have never been observed to spontaneously select
a microscopic state able to sustain the applied stress, while, conversely, in the presence
of friction a much more complex phenomenology has been found~\cite{Zhang,Makse,van
Hecke,Magnanimo,Behringer,Grebenkow,Sandnes,PRE2011}. 
In particular, in recent works~\cite{Grebenkow,PRE2011} reporting molecular dynamics (MD)
simulations of soft frictional grains at constant volume and applied shear stress, 
it was observed the phenomenology who inspired the idea of 'fragile matter'~\cite{Farr}. Indeed,
for some values of the control parameters, 
the system was found to select a configuration able to sustain the applied stress 
after a small slip, or even after very large strains at constant rate~\cite{phil mag}. 

Under these conditions, whether the mechanical response of the system 
is peculiar or rather similar to that of
other amorphous and soft materials~\cite{Goldhirsch2}
is still an open question.  
In particular it is not clear to what extent
fragility can capture the physics of these systems. In more concrete terms, can
small incompatible stresses restore the flow?  

In this paper, we investigate via MD simulations the mechanical response of frictional granular
systems jammed in the presence of a shear stress.
After shortly reviewing the numerical model and the overall phenomenology (Sec. \ref{Sys}),
we discuss the limit of validity of the concept of `fragile matter' (Sec. \ref{sh_m}).
We find that our system is not fragile but it responds as an elastic and isotropic solid
to small incompatible stresses. Anisotropy and macroscopic rearrangements
only emerge in the response to strong stress variations.
In Sec. \ref{str-cyc}, we will focus on the stress-strain curves during
a stress cycle, in order to check analogies with
ordinary rheological behaviours. 
Increasing the maximum stress reached during the cycle, we find
a crossover from a visco-elastic to a plastic regime.

\section{Investigated System}
\label{Sys}
We perform MD simulations along the line of Refs~\cite{Grebenkow,PRE2011,phil mag}.
Monodisperse spherical grains of mass $m$ and diameter $D$ are enclosed in a box
of dimension $l_x = l_y = 16D$, and $l_{z} = 8\textit{D}$. Periodic boundary
conditions are used along $x$ and $y$, while the size of the vertical dimension
is fixed and chosen to be comparable to that of recent experiments~\cite{Pine,Daniels}.
The upper and lower boundary surfaces of the box are disordered collections of particles
that move as a rigid object. The bottom plate has an infinite mass, and is
therefore fixed, while the top one has a mass equal to the sum of the masses of its particles
(roughly $l_xl_y$). We impose a shear stress $\sigma_{zx} = \sigma$
to the system, applying a force to the top plate.

Grains interact via the standard linear spring-dashpot model.  Two particles $i$
and
$j$, in positions ${\bf r}_{i}$ and ${\bf r}_{j}$, with linear velocities ${\bf
v}_{i}$ and ${\bf v}_{j}$,
and angular velocities $\omega_{i}$ and $\omega_{j}$, interact if in contact,
i.e., if the quantity
$\delta_{ij}=D-|{\bf r}_{ij}|$ is positive. $\delta_{ij}$ is called the
penetration length, and ${\bf r}_{ij}={\bf r}_{i}-{\bf r}_{j}$ is the distance
between particles $i$ and $j$. The interaction force has a
normal component ${\bf F}_{n_{ij}}$  and a tangential one ${\bf F}_{t_{ij}}$:

\begin{equation}
\label{eq1}
 {\bf F}_{n_{ij}} = -k_{n} \delta_{ij} {\bf n}_{ij} - \gamma_n m_{eff} {\bf
v}_{{\bf n}_{ij}}
\end{equation}

\begin{equation}
\label{eq2}
 {\bf F}_{t_{ij}} = -k_{t} {\bf u}_{t_{ij}} .
\end{equation}

Here $k_n$ and $k_{t}$ are elastic moduli, 
${\bf n}_{ij} = {\bf r}_{ij}/|{\bf r}_{ij}|$, ${\bf v}_{n_{ij}} =
[({\bf v}_{i} - {\bf v}_{j}) \cdot {\bf n_{ij}}] {\bf n}_{ij}$, ${\bf v}_{{\bf
t}_{ij}} = {\bf v}_{{ij}} - {\bf v}_{{\bf n}_{ij}}$, $m_{eff}$ is the reduced
mass, and $\gamma_n$ accounts for dissipative character of the normal component.
${\bf u}_{\bf t_{ij}}$, set to zero at the beginning of a contact, measures the
shear displacement during the lifetime of a contact. Its time evolution is fixed
by ${\bf v}_{{\bf t}_{ij}}$  and $\omega_{i}$ and $\omega_{j}$, as
described in Ref.~\cite{Silbert}. The presence of tangential forces implies the
presence of torques, $\tau_{ij} = -1/2 {\bf r}_{ij}\times {\bf F}_{t_{ij}}$.
We use the value of the parameters of~\cite{Silbert}: $k_n=2~10^{5}$,
$k_{t}/k_{n}=2/7$, $\gamma_{n}=50$. 
Length, masses and times are expressed in units of $D$, $m$ and $\sqrt{m/k_n}$.
We vary the volume fraction $\phi$, which represents the volume occupied by the grains
divided by the volume of the container, by changing the number of particles.
The initial state is prepared setting to zero the friction coefficient~\cite{Song}, 
randomly placing small particles into the system, and then inflating them until the
desired volume fraction is obtained; such a protocol is a short-cut of
experimental procedures with which it is possible to generate very dense
disordered states of frictional systems, such
as oscillations of high frequency and small amplitude~\cite{Gao}.

\begin{figure}[t!]
\begin{center}
\includegraphics[width=0.3\textwidth]{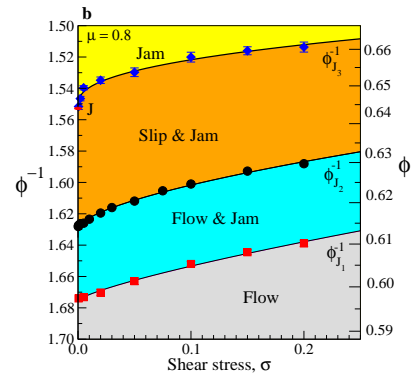}
\end{center}
\caption{(Color online) Jamming phase diagram in the density-shear stress
 ($\phi-\sigma$) space at a fixed value of the friction, $\mu=0.8$. Three transition lines bound the
 different regions.
 } 
\label{diagram}       
\end{figure}

We have recently investigated the jamming properties
of this system~\cite{PRE2011} as $\phi$, $\sigma$, and $\mu$ are varied, and 
summarize our findings below.
Confirming previous results~\cite{PCC09},
in the frictionless case the system is either found in a flow or in a jammed phase, a
single line marking the transition between these two phase in the $\phi$--$\sigma$ ($\mu = 0$) plane.
By contrasts, at a finite value of the friction coefficient, more
complex rheological regimes are found, corresponding to four distinct regions in the $\phi$--$\sigma$ ($\mu > 0$) plane, as
illustrated in the schematic jamming phase diagram reported in Fig.~\ref{diagram}. 

\begin{figure}[t!]
\begin{center}
\includegraphics*[width=0.4\textwidth]{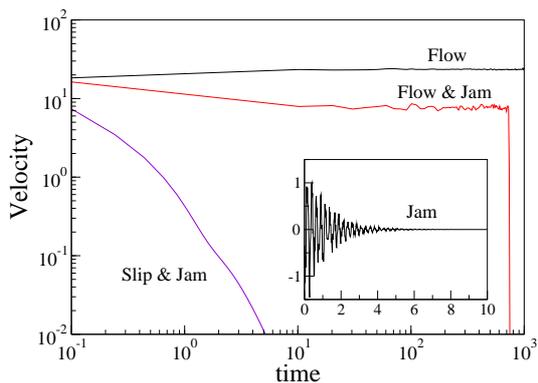}
\end{center}
\caption {Typical behaviour of the velocity of the top plate as a function of time in the different regimes, from Ref. \cite{PRE2011}. Here $\sigma=0.5$, $\mu=0.8$ and $\phi=0.578$ (Flow), $\phi=0.596$ (Flow \& Jam),  $\phi=0.629$ (Slip \& Jam), $\phi=0.655$ (Jam)}
\label{velocity}
\end{figure}

On increasing the volume fraction the system visits the following regimes: Flow, Flow \& Jam, Slip \& Jam, Jam.
In each of these regime the system exhibits a different response to the applied
shear stress $\sigma$, as illustrated in Fig.~\ref{velocity} where we show the time
evolution of the velocity of the shearing top plate. The 
different regimes can be characterized as follows: \\
\noindent {\bf Flow:} the system flows with a steady velocity reached after a
transient.\\ 
\noindent {\bf Flow \& Jam:} the system reaches a steady velocity after a
transient. However, after flowing for a time, $t_{jam}$, it suddenly jams.\\ 
\noindent {\bf Slip \& Jam:} the system jams
after a small inelastic displacement of the top plate. Steady flow is never observed. \\ 
 \noindent {\bf Jam:} the system responds as an elastic solid.

The value of the volume fraction ($\phi_{j_{1}}$, $\phi_{j_{2}}$ and $\phi_{j_{3}}$) \cite{comm1}
marking the transition between the
different regimes depend both on $\sigma$ and $\mu$, and their identification~\cite{PRE2011} 
allow to describe the overall phenomenology
in a three-dimensional jamming phase diagram in the $\phi, \sigma, \mu$ space. In the limit
$\sigma \to 0$ the transition from the `Flow \& Jam' to the `Jam' regime appears
to occur at a volume fraction identifiable with random--close packing volume fraction,
if one neglects its protocol dependence~\cite{PC2010}. We prefer not to associate the other transitions
volume fraction to the random loose volume fraction, considering that this has been introduced
using a different protocol (pouring), and that this is also expected to be protocol dependent~\cite{RVLP}.

We have previously investigated the dynamical and geometrical properties
in the different regimes~\cite{PRE2011}. For example, we find that
in the `Flow \& Jam' region the average value of $t_{jam}$ diverges on approaching 
the `Flow' regime, thus allowing to identify the transition line $\phi_{j_{1}}$ between `Flow' and `Flow \& Jam' region. 
In addition, the $t_{jam}$ distribution is found to be extremely broad.
Moreover, for each value of the friction, $\mu$, we find the mean contact number
$Z$ to be constant across the `Flow \& Jam' and the `Slip \& Jam' regimes.
This is in agreement with Ref.~\cite{Song}, even though we also found the constant
value of $Z$ to depend on the applied shear stress. We described in Ref.~\cite{phil mag}
the rheology of the system and a possible mechanism for the transition from a flowing to a jammed state.

In this paper, we fix the Coulomb friction coefficient to~$\mu=0.8$, and consider values
of $\phi$ and $\sigma$ corresponding to the `Flow \& Jam' and to the `Slip \& Jam' regime,
where the systems jams after flowing, and could therefore be fragile. 
In the `Jam' region fragility is not expected  
as jamming is not preceded by a transient flow.  

\begin{figure}[t!]
\begin{center}
\includegraphics[width=0.33\textwidth]{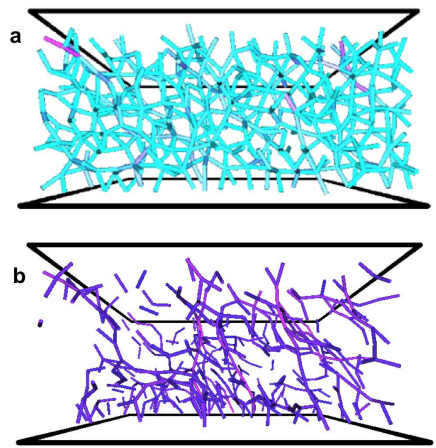}
\end{center}
\caption{(Color online) a) Network of contact forces in a system jammed under the effect of a constant shear stress. Each segment marks the
direction of the force acting between a pair of grains in contact. A color scale is used to represent the force intensities. b) The picture
only shows the strongest contacts ($10\%$ of all contacts) of the network. } 
\label{perc}       
\end{figure}

\section{Mechanical response of jammed states}
\label{sh_m}

A minimal model for the structure of fragile systems consists
in a series of force chains directed along the strongest stress direction, 
supporting the applied stress and living in a sea of spectators~\cite{Cates1,Cates2,Cates3}.
We show the network of all forces of a system jammed under the action of a shear
stress in Fig.~\ref{perc}{\bf a}. 
A percolating cluster of contacting particles furnishes the support for the network of force.
Finite clusters, except few single particles, called rattlers, are not allowed, 
since, due to repulsive force acting between
particles in contact, a cluster not hold by the confining plates breaks.
The network may be also explored investigating
how it changes as a function of a threshold on the inter--particle force one may
introduce to define the bonds. For instance, in Fig.~\ref{perc}{\bf b} 
we show the network of the strongest forces, obtained using a 
threshold which allow to pick--up roughly $10\%$ of all forces (strongest ones).
While the network of Fig.~\ref{perc}{\bf a} appears to be isotropic, that of Fig.~\ref{perc}{\bf b}
appears  highly anisotropic, with force chains lying in the direction of the strongest stress,
and resembles that mentioned by Cates et al.~\cite{Cates1,Cates2,Cates3}.
If the response to external perturbations is dominated by these strong forces, one may
therefore expected a `fragile' behavior.

In order to check this point,
we probed the elastic properties of a system jammed under the action of the existing
shear stress, $\sigma_{zx}=\sigma$, by superimposing a perturbing shear stress.
The non-zero components of this perturbing stress are $\delta \sigma_{zx}$ and $\delta
\sigma_{zy}$, we fix in such a way that $\delta \sigma_{zx}^2 + \delta \sigma_{zy}^2 =
\delta\sigma^2$. The perturbing shear stress is therefore conveniently expressed
in terms of $\delta \sigma$ and of $\theta = \arctan\left(\delta
\sigma_{zy}/\delta \sigma_{zx}\right)$.

\begin{figure}[t!]
\begin{center}
\includegraphics*[width=0.5\textwidth]{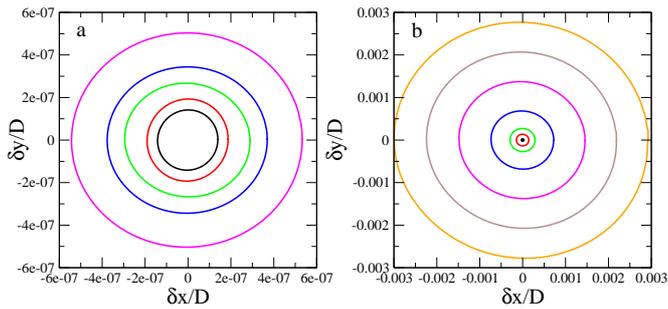}
\end{center}
\caption{ Response of a jammed system to a small perturbing shear stress. Panel (a) shows the response
at $\sigma = 10^{-2}$, and $\delta \sigma = 10^{-4}$ for different values of the volume fraction (from the inside, $\phi =$
$0.655$, $0.630$, $0.617$, $0.613$ and $0.610$). Panel (b) shows the response at $\phi = 0.617$, $\sigma = 10^{-2}$,
for different values of the perturbing stress (from the inside, $\delta \sigma = $ $10^{-3}$, $5~10^{-3}$, $10^{-2}$, $2.5~10^{-2}$,
$5~10^{-2}$,
$7.5~10^{-2}$, $10^{-1}$).}
\label{circles}
\end{figure}

Figure~\ref{circles} shows the displacement ${\bf \delta r} = (\delta x,\delta y)$ of the top plate
position for different values of the volume fraction at fixed $\sigma$ and $\delta\sigma$ (left),
and for different values of $\delta \sigma$ at fixed $\sigma$ and $\phi$ (right).
Each curve is obtained applying a perturbing shear stress with ($\theta
= 0$), and then increasing $\theta$ from $0$ to $2\pi$.
While open path were expected for restored flow,
we find that each curve describes a close path, which implies that the system responds elastically to the applied force. 
Moreover, this path is to a good approximation a circle ($ |{\bf \delta r}| \simeq const$), 
which implies that the elastic response is the
same for all values of $\theta$.
An estimation of the degree of anisotropy in the response if obtained by measuring the
parameter
\begin{equation}
 \xi(\theta) = \frac{\left[ \delta x^2 (\theta) + \delta y^2 (\theta)
\right]^{1/2} - \overline{\delta r} }{\overline{\delta r}},
\end{equation}
where $\overline{\delta r} = \langle \left[ \delta x^2 (\theta) + \delta y^2
(\theta) \right]^{1/2} \rangle_\theta$.

\begin{figure}[t!]
\begin{center}
\includegraphics*[width=0.35\textwidth]{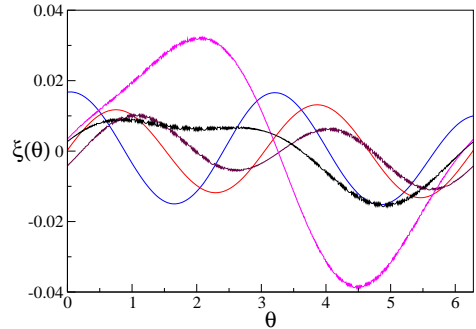}
\end{center}
\caption {(Color online)Anisotropy in the response to a small perturbing shear stress
of a system jammed under the action of a large shear stress.
Different curves refer to different values of the volume fraction.}
\label{anisotropy}
\end{figure}

As illustrated in Fig. \ref{anisotropy} , the anisotropy of the system is small, being $|\xi(\theta)| < 4\%$ and does not reveal any
particular pattern.

Being the response  to small $\delta\sigma$ 
elastic (the strain is proportional to the stress) and to a good approximation
isotropic (the strain does not depend on $\theta$),
as clarified by Fig. \ref{circles} and Fig. \ref{anisotropy},  
the system is characterized by a well defined 
shear modulus $G = \lim_{\delta \sigma \to 0} \delta
\sigma/\epsilon$, where $\epsilon={\bf \delta r}/L_{z}$ is the
shear strain induced by $\delta \sigma$. 
As a further characterization of the mechanical properties of the system, we
have studied the Hessian along the line of Ref.~\cite{Somfai},
only finding zero eigenvalues (expected due to the presence of
rattlers), and negative ones, while positive values are expected
for fragile solids. 

This analysis clarify that frictional granular systems
jammed at constant volume and applied shear stress 
are not fragile, as they respond elastically to small
perturbations, regardless to their orientation.
\begin{figure}
\begin{center}
\includegraphics* [width=0.25\textwidth]{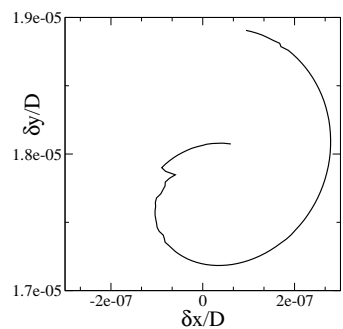}
\end{center}
\caption{Using the same protocol of Fig. \ref{circles}, response of a jammed system at $\phi=0.606$, $\sigma=10^{-2}$ to a perturbation $\delta \sigma=10^{-4}$. The open path signals the presence of restored flow.} 
\label{spiral}       
\end{figure}
The absence of fragility can be rationalized in terms of
the properties of the energy landscape of the system. Indeed,
fragile jammed systems can be associated with saddle points,
as their elastic energy may increase or decrease, depending
on the direction of the perturbation, respectively leading to an
elastic response or to an instability. Since dissipative systems
do not spontaneously arrest in an unstable point of their energy
landscape, we expect them to arrest in a true energy minimum.
Systems that jam under the action of an applied stress are,
therefore, not expected to be fragile. Of course, restored flow predicted by fragility,
may appear in response to large stress variations, which are able
to carry the system away from an energy minimum.
Indeed, at higher $\delta \sigma$, we find that curves like those in Fig. \ref{circles} become more and
more elliptic, signaling the emergence of anisotropy. At a threshold $\delta \sigma_{c}$
these curves turn in open paths, such as
the spiral shown in Fig. \ref{spiral}, which evidence the presence of restored flow.
The value of the threshold depends on the volume fraction, 
$\delta \sigma_{c}$=$\delta \sigma_{c}(\phi)$ as well as on the applied shear stress.
In the explored rage of volume fractions, we have never observed a fragile behavior when the relative variation
of the shear stress is $\delta \sigma_{c}/\sigma < 10^{-3}$.

These results suggest that in the response to small perturbations
the systems probes the roughly isotropic network of all forces of Fig.~\ref{perc}a,
rather than the strong network of Fig.~\ref{perc}b.

\section {Stress cycle}
\label{str-cyc}
\begin{figure}[t!]
\begin{center}
\includegraphics*[width=0.4\textwidth]{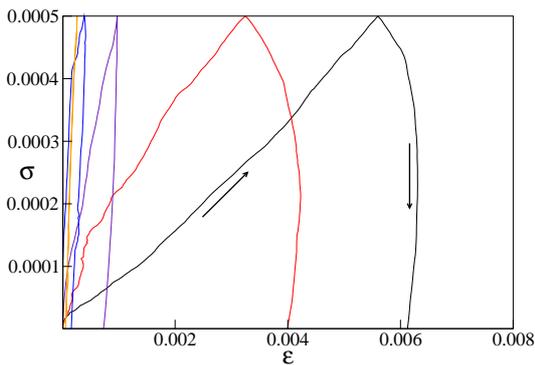}
\end{center}
\caption{ (Color online) Displacement of the top plate in a stress cycle. The
stress is first increased to its maximum value $\sigma_{m}$ and then decreased to
zero. Here $\sigma_{m} = 5~10^{-4}$ and $\mu = 0.8$.
From right to left, $\phi = 0.6475$, $0.6477$, $0.6480$, $0.6482$, $0.6488$.}
\label{e_phi}
\end{figure}
As discussed in the previous section, the mechanical
response of jammed grains in the investigated range of
control parameters resembles that of other elastic solids. 
However, granular systems are also expected to 
exhibit a plastic response, as other disordered systems.
We have investigated this possibility probing the response of the system
to stress-strain cycles.
In~\cite{PRE2011} we have used this approach to
locate the line separating
the `Slip \& Jam' and the `Jam' region 
in the jamming phase diagram of Fig. \ref{diagram}. Here we focus
on the degree of plasticity emerging in the solid response
for increasing value of the shear stress.

We have measured stress-strain curves 
when the system undergoes a stress-cycle for values of the control
parameters which span from the `Flow \& Jam' to the the `Jam' region. 
Precisely, after preparing the system, we slowly increase the shear stress $\sigma$
up to a maximum value $\sigma_{m}$, and then decrease it to zero. After
each stress cycle we measure the residual strain $\epsilon=\Delta L
/L_{z}$, where $\Delta L$ is the displacement of the top plate.

\begin{figure}[t!]
\begin{center}
\includegraphics*[width=0.38\textwidth]{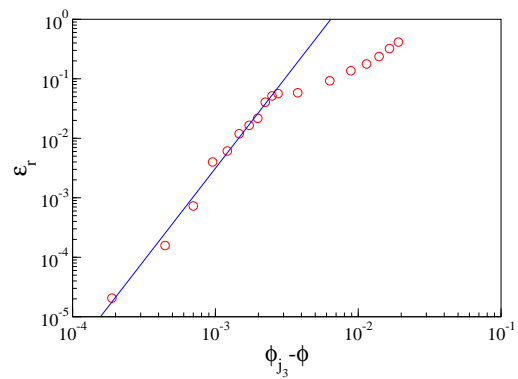}
\end{center}
\caption{ (Color online) For a fixed value of the shear stress ($\sigma_{m} = 5~10^{-3}$), 
the residual strain $\epsilon_{r}$ decreases on increasing the volume
fraction, and vanishes at a volume fraction $\phi_{j_{3}}$, which depends on $\sigma_{m}$ and
$\mu$. The straight line is a power law $\epsilon_{r} = a (\phi_{j_{3}}-\phi)^b$, $b \simeq 1.2$ and $\phi_{j_{3}} \simeq 0.6495$.}
\label {fi_c}
\end{figure}

\begin{figure}[t!]
\begin{center}
\includegraphics*[width=0.4\textwidth]{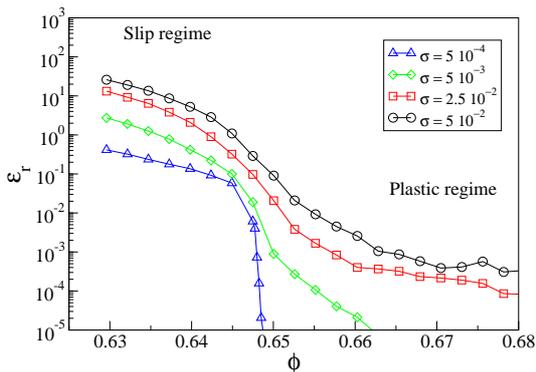}
\end{center}
\caption{(Color online) Dependence of the residual strain $\epsilon_{r}$ on the volume fraction $\phi$, for different
values of $\sigma_{m}$, at $\mu = 0.8$.}
\label{crossover}
\end{figure}

At low volume fraction, in the `Flow \& Jam' regime, the initial state of
the system is not jammed, and accordingly we expect a finite residual
strain at the end of the cycle, $\epsilon_{r} > 0$. Conversely, at higher volume fraction
the system has a solid response, we expect to be elastic ($\epsilon_{r} =0$)
at small $\sigma_m$, and plastic ($\epsilon_{r} >0$) for  $\sigma_m$ overcoming the yield stress.
Figure~\ref{e_phi} shows the strain as a function of the shear
stress for a small value of the maximum stress,  
$\sigma_{m}=10^{-4}$, and different volume fractions. 
Accordingly to the expectations,  
the residual strain becomes smaller as the density increases,
and appears to critically vanish at
$\phi_{j_{3}}$, which has been determined via a numerical fit, as shown
in Fig.~\ref{fi_c}. On increasing $\sigma_m$, a finite
residual strain $\epsilon_{r} > 0$ is also found at higher volume fractions.
This is shown in Fig.~\ref{crossover}, where we observe that $\epsilon_{r}$ vanishes on increasing
$\phi$ at small $\sigma_m$, while it decreases and then bends at larger $\sigma_m$.

The crossover in the behavior of $\epsilon_r$ can be related to a change from a visco--elastic to a plastic regime,
and may be explained focusing on the increase of the number of contacts that break as the strain
increases. At small stress, the strain of the system is small,
contacts do not break and the system responds elastically. At higher stress, the strain of the
system is large, and contacts break. This is the microscopic
origin of the plastic response, as the system looses memory of the
tangential force of the contacts reaching the Coulomb threshold.

\section{Conclusion}
\label{conclusion} 
We have investigated  the mechanical response to small perturbations
of systems jammed under the action of an applied shear stress,
and found this to be elastic and isotropic, and therefore not
consistent with the expectation of a `fragile' state. This result is explained considering that 
the response to small perturbations see the cooperation of all contact forces, which are isotropically
distributed in the systems. Conversely, the response to large perturbation is due to the strongest forces,
which are not isotropic. Accordingly, we have found that large perturbations
lead either to an anisotropic response, or to unjamming.
The investigation of the response of the system to stress--cycle, also allowed
to observe the transition from an elastic to a plastic response as the maximum applied stress
increases. 

\begin{acknowledgements}
We acknowledge computer resources from the University of Naples Scope grid project, CINECA, CASPUR and DEISA.
\end{acknowledgements}

\bibliographystyle{spphys}       


\end{document}